\documentclass[12pt,preprint]{aastex}
\usepackage{graphicx}

\newcommand{\astroh}{{\small \it Astro-H}}

\newcommand{\agn}{{\small AGN}}
\newcommand{\ngc}{{\small NGC\,3783}}
\newcommand{\ngcseven}{{\small NGC\,7469}}
\newcommand{\ngcfifty}{{\small NGC\,5548}}
\newcommand{\mcg}{{\small MCG\,--6--30--15}}
\newcommand{\iras}{{\small IRAS~13349+2438}}

\newcommand{\xstar}{{\small XSTAR}}

\newcommand{\xmm}{{\it XMM-Newton}}
\newcommand{\rgs}{{\small RGS}}
\newcommand{\chandra}{{\it Chandra}}
\newcommand{\letg}{{\small LETG}}
\newcommand{\hetg}{{\small HETGS}}

\newcommand{\x}{X-ray}

\newcommand{\kms}{km~s$^{-1}$}
\newcommand{\xicgs}{erg\,s$^{-1}$\,cm}
\def\lsim{\mathrel{\rlap{\lower4pt\hbox{$\sim$}}
    \raise1pt\hbox{$<$}}}                
\def\gsim{\mathrel{\rlap{\lower4pt\hbox{$\sim$}}
    \raise1pt\hbox{$>$}}}                

\begin{document}

\title{Density Profiles in Seyfert Outflows}
\author{ Ehud Behar \altaffilmark{1, 2}
}

\altaffiltext{1}{Senior NPP Fellow, NASA / Goddard Space Flight Center, Greenbelt MD 20771. \\
		behar@milkyway.gsfc.nasa.gov}
\altaffiltext{2}{Permanent address: Department of Physics,
                 Technion, Haifa 32000, Israel. 
}

\received{} \revised{} \accepted{}

\shorttitle{Density Profiles in Seyfert Outflows}
\shortauthors{Behar}

\begin{abstract}
For the past decade, ionized outflows of a few 100~\kms\ from nearby Seyfert galaxies have been studied in great detail using high resolution X-ray absorption spectra.
A recurring feature of these outflows is their broad ionization distribution including essentially ions (e.g., of Fe) from neutral to fully ionized.
The absorption measure distribution ($AMD$) is defined as the distribution of column density with ionization parameter $\left| d N_H/d (\log \xi) \right|$.
$AMD$s of Seyfert outflows can span up to five orders of magnitude in $\xi$.
We present the $AMD$ of five outflows and show that they are all rather flat, perhaps slightly rising towards high ionization.
More quantitatively, a power-law fit for $\log AMD \propto (\log \xi )^a$ yields slopes of $0 < a < 0.4$.
These slopes tightly constrain the density profiles of the wind, which until now could be addressed only by theory.
If the wind is distributed on large scales, the measured slopes imply a generic density radial profile of $n \propto r^{- \alpha}$ with $1 < \alpha < 1.3$.
This scaling rules out a mass conserving radial flow of $n \propto r^{-2}$, or a constant density absorber, but is consistent with a non-spherical MHD outflow model in which $n \propto r^{-1}$ along any given line of sight.
On the other hand, if ionization variations are a result of local ($\delta r$) density gradients, e.g. as in the turbulent interstellar medium (ISM), the $AMD$ slopes imply density scaling of $n \propto \delta r^{- \alpha}$ with $0.7 < \alpha < 1.0$, which is quite different from the scaling of approximately $n \propto \delta r^{0.4}$ found in the Milky Way ISM and typical of incompressible turbulence. 

\end{abstract}


\section{INTRODUCTION}
\label{sec:intro}

The \x\ spectra of many active galactic nuclei (\agn s) viewed
directly toward the central source (e.g., Seyfert 1 galaxies) show
the continuum flux absorbed by numerous absorption lines produced
by ionized gas.
The role of these absorbers for the \agn\ and for the host galaxy are still being debated.
In order to advance our understanding of these outflows and their significance to the \agn, it is important to develop model-independent techniques for quantifying the physical parameters of the wind.
So far, only the wind velocity measured from Doppler shifts and the total
column density $N_H$ deduced from the equivalent widths of absorption 
lines are known with high confidence.
The broad range of ionization (and temperature) in the wind is also unambiguous.
Conversely, there is little robust constraints on the location, geometry, and density 
of the wind, which leaves its mass outflow rate, momentum, and energy fairly uncertain. 

The \x\ band can potentially provide the full physical picture of \agn~outflows,
as it comprises detectable absorption lines from the
full range of charge states: From neutral up to hydrogen-like.
The detection of all of the absorbing charge states of a given element 
enables an accurate measurement of the total column density.
For this purpose, Fe ions are particularly useful as they form over 
several orders of magnitude of ionization parameter $\xi$

\begin{equation}
\xi=\frac{L}{nr^2}
\label{xi}
\end{equation}

\noindent where $L$ is the ionizing luminosity between 1 and 1000
Rydberg, $n$ denotes the hydrogen number density and $r$ is the
distance of the absorber from the ionizing source.  
\agn\ outflows can be distributed over the full range of ionization $-1 < \log \xi < 4$
\citep{steenbrugge03, steenbrugge05, costantini07, tomer07}.
In order to best quantify this distribution, \citet{tomer07} defined an absorption 
measure distribution ($AMD$), which is the distribution of the hydrogen column density 
$N_H$ along the line of sight as a function of $\log \xi$

\begin{equation}
AMD=\left| d N_H/ d (\log \xi) \right|
\label{AMD}
\end{equation}

\noindent Reciprocally, the total $N_H$ can be expressed as an integral over the $AMD$.
The $AMD$ is the absorption analog of the Emission Measure
Distribution ($EMD$) widely used in the analysis of emission-line spectra.
It provides a more complete representation of the ionization 
distribution than the more commonly used models of several ionization components
each with a fixed $\xi$; the overall success of the latter in producing good spectral fits to the data
notwithstanding. 
The $AMD$ reconstruction method is outlined in detail in \citet{tomer07}. 
Here, we only note that for each target,
the observed continuum spectrum is used to obtain the ionization balance as a function of $\xi$ from the \xstar\ code \citep{kallman01}.
Subsequently, the $AMD$ is obtained by a fitting process that aims to best reproduce all of the individually measured ionic column densities.
Being the solution of an inversion problem, the reconstructed $AMD$ is inevitably degenerate.
Nonetheless, a step function is the preferred $AMD$ form, as
it allows rigorous local error calculations that take into account the degeneracy between 
different $\xi$ bins. For more details see \citet{tomer07}.

In this paper, we present the available $AMD$ distributions,
which were derived for five sources with high signal-to-noise-ratio \x\ grating spectra.
The objective is not to focus on any individual target,
but to seek trends in the best available (yet admittedly small and incomplete) sample of \agn\ outflows.
The idea is to link the general $AMD$ behavior with the physical properties of 
the outflows, in particular the density profile, in order to obtain an insight into the 
origin and astrophysical significance of the winds. 
These physical implications are described in \S \ref{phys} and discussed in \S \ref{discussion}

\section{Targets and $AMD$s}
\label{sec:data}

We include all the Seyfert outflow targets with sufficiently good grating spectra for which we were able to reconstruct a non-degenerate $AMD$.
The $AMD$s of \iras\  and of \ngc\ were presented in great detail in \citet{tomer07}
and are based on deep exposures with the \hetg\ spectrometer on board \chandra.
The $AMD$ of \mcg\ is from a recent analysis of the \hetg\ spectrum (Holczer et~al., in preparation) and corresponds to the slow \agn\ component in that complex spectrum.
The $AMD$ of \ngcfifty\ was derived from the ionic column densities published by \citet{steenbrugge05} and based on spectra from the \hetg\ and \letg\ instruments on board \chandra.
The $AMD$ of \ngcseven\ is an improved version of that presented in \citet{blustin07}
and is based on spectra acquired with the \rgs\ spectrometer on board \xmm.
The improvement is that unlike in \citet{blustin07}, here we followed the general method of \citet{tomer07} and allowed the $AMD$ to be zero in certain bins, if required by the fit.
We note that a thorough spectral analysis was carried out also for the \chandra\ \letg\ spectrum of Mrk~279 by \citet{costantini07}.
However, the low column density and marginal detection of Fe absorption in Mrk~279 is insufficient to produce a well constrained $AMD$.

\begin{figure}
\centerline{\includegraphics[width=13cm,angle=-90]{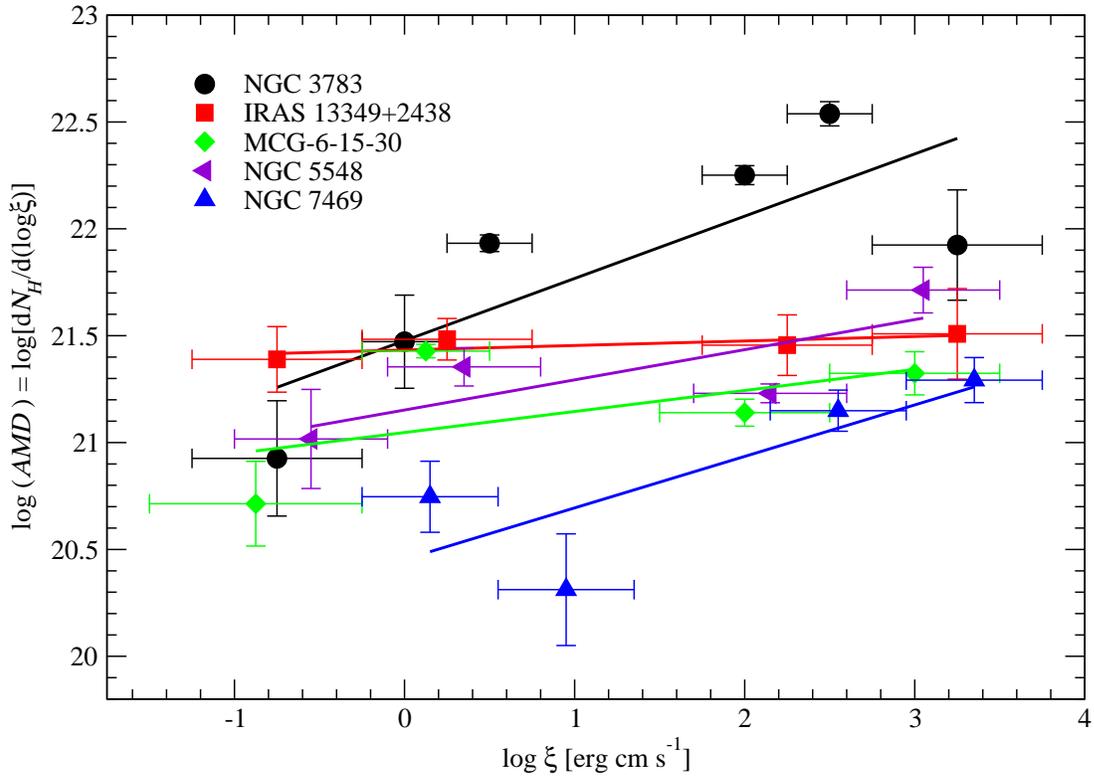}}

  \caption{$AMD$ distributions obtained for five targets. 
  The best fit power-law is presented for each $AMD$. 
  The inferred power-law slopes are relatively moderate, indicating a rather
  flat distribution of column density with ionization parameter.
   }
   \label{fAMD}
\end{figure}

The five $AMD$s used in the present work are plotted in Fig.~\ref{fAMD}.
Note that all $AMD$s have a gap at intermediate $\xi$ values,
which is in fact a data point at zero that is not plotted.
For four sources this gap falls between $\log \xi \approx 0.5~- 1.5$~\xicgs, while for \ngcseven\ it is at slightly higher values.
In \citet{tomer07} and in \citet{blustin07}, these $AMD$ discontinuities were interpreted as
regions of thermal instability.
A physically separated two-component outflow is another possible interpretation,
although the similar outflow velocities of the low and high ionization regions  
and the ubiquity of the minimum at approximately
the same temperature regime argue against two separate components.
The presence of the thermally unstable region and its exact position (in $\xi$) 
depend strongly on the correspondence between charge state and $\xi$ that are derived from \xstar\ \citep{kallman01} and are uncertain to some degree.
It is known, for instance, that some of the available dielectronic recombination (DR) rates used in these computations are inadequate \citep{netzer04, badnell06}.
Another effect that could potentially change the unstable region by affecting the 
$\xi$ at which low charge states form in the outflow is photoionization of these ions by the EUV continuum that is poorly constrained.
Finally, all ionization balance computations are carried out with a single ionizing spectrum, i.e., in the optically thin limit.
This is justified by the observed \x\ spectra that typically show only shallow photoelectric edges, if any.
Once the medium becomes optically thick and deeper regions are exposed to lower (and harder) ionizing luminosity, this approximation gradually loses its validity. 

In this work, we are interested more in the broad characteristics of the distribution.
Assuming that the minimum is due to thermal instability justifies viewing the remaining
stable regions of the $AMD$ as different ionization components of the same outflow that are likely driven by the same physical mechanisms.
We thus apply simple (least mean square) linear fits to each $AMD$ according to 
 $\log AMD (\xi) = a\log \xi + b$, or $AMD (\xi)= 10^b\xi ^a$.
The allegedly thermally unstable ionization regions are ignored in the fit.
A similar power law distribution of ionization was suggested by \citet{steenbrugge03}.
It can be seen in Fig.~\ref{fAMD} that a power law representation of the $AMD$ for some targets is far from perfect.
Nevertheless, given the uncertainties on each data point, and for the mere purpose of seeking $AMD$ trends, we deem it sufficient.
Moreover, a power law $AMD$ representation allows direct comparisons with self-similar wind models as discussed in the following sections.
The best-fit power laws for the five outflows are shown in Fig.~\ref{fAMD} and the fit parameters are listed in Table~\ref{Tfit}.
Evidently, all five power laws have relatively similar slopes of $a = 0.0 - 0.4$. 
The implied density profile slopes ($\alpha$), introduced in the next section, are also listed in Table~\ref{Tfit}.
\citet{steenbrugge05} fitted a power law to their distribution of individual ionic column densities in \ngcfifty, which is somewhat similar but not identical to the full $AMD$ \citep[see][for details]{tomer07}.
They obtained a slope of $a = 0.2 \pm 0.1$ (denoted there $\alpha$) with Fe ions alone, which is in excellent agreement with our best-fit of $a = 0.14 \pm 0.08$.
When they use other elements, however, and need to rely on the assumption of solar abundances,
 they get a somewhat steeper slope of  $a = 0.4$.
 
\section{Physical Implications for Outflows}
\label{phys}

In this section, we explore what the $AMD$ shape, namely its slope, implies for the density profile of the outflow.
It is convenient for this purpose to consider two separate boundary cases.
One is a radial large-scale outflow in which the density $n(r)$ varies as a function of the distance from the source $r$.
In this case, $\xi = L/(nr^2)$ varies along the flow due to the combined change in $r$ and in $n$.
The other possibility is a remote absorber at a distance more or less constant $r = r_0$,
but with strong density gradients over small distances $\delta r \ll r_0$.
In this case, the ionization varies exclusively due to the local density gradients. 
Obviously, a real outflow could be a combination of the two, namely a large scale outflow with strong local density gradients.
However, for the current analysis, it is instructive to consider these two cases separately.

\subsection{Large Scale Outflow}
\label{large}

A large-scale outflow is defined here as an outflow with a size comparable to the distance $r$ from the ionizing source so that the ionization parameter (eq. \ref{xi}) is a function of $r$ also through the parameterization of $n(r)$.  Let us assume a power-law density profile with an index $\alpha > 0$

\begin{equation}
n(r) \propto r^{-\alpha}
\label{nalpha}
\end{equation}

%
\noindent Such a parameterization of the density is convenient for comparison with self-similar wind solutions. The ionization parameter thus varies with radius as $\xi (r) \propto r^{\alpha -2}$, or reciprocally $r (\xi) \propto \xi ^{1/(\alpha - 2)}$
and

\begin{equation}
\frac{d r}{d \xi} \propto \xi ^{-\frac{\alpha - 3}{\alpha - 2}} 
\label{xir}
\end{equation}

\noindent The radial increment of hydrogen column density $N_H$ with the ionization parameter $\xi (r)$, can be written as

\begin{equation}
dN_H \equiv n(r)dr = n(r)\frac{d r}{d \xi}d\xi \propto 
\xi ^{-\frac{2\alpha -3}{\alpha - 2}} d\xi
\label{dnh}
\end{equation}

\noindent Consequently, the $AMD$ (eq. \ref{AMD}) scales with $\xi$ as 

\begin{equation}
AMD = \left| \frac {d N_H}{ d \log \xi} \right| 
\propto \xi \left| \frac {d N_H}{d \xi} \right| \propto \xi ^{-\frac{\alpha - 1}{\alpha - 2}} \equiv \xi ^a 
\label{AMDxi}
\end{equation}

\noindent 
where $a= -(\alpha - 1)/(\alpha - 2)$ is precisely the slope of $\log AMD$ vs. $\log \xi$ fitted for in Fig.~\ref{fAMD}.
Finally, $\alpha$ and its uncertainty can be expresses as

\begin{equation}
\alpha = \frac{1+2a}{1+a}   \pm \frac{\Delta a}{\left( 1+a \right)^2} 
\label{alpha}
\end{equation}

\noindent where  $\Delta a$ stands for the standard error of the fitted power law slope $a$.
The resulting $\alpha$ values for the five outflows are listed in Table~\ref{Tfit}.
Note that a scaling with $\xi$ similar to that of eq. (\ref{AMDxi}) is obtained for the integrated $N_H$ up to $r$, as $N_H(r) \propto \int_0^r n(r')dr' \propto r^{1-\alpha} \propto \xi ^{-\frac{\alpha - 1}{\alpha - 2}}$. 
However, $N_H(r)$ is a property that pertains to the entire flow up to radius $r$,
while the $AMD$ is a local property of the flow at each position (or as a function of $\xi$),
 and is therefore preferred here.

Note that for $AMD$s that increase with $\xi$ (i.e., flow dominated by high charge states) as found in the present sample, $\alpha$ must obtain values between 1 -- 2.
The observed slopes, e.g., of $a =$ 0, 0.25, or 0.4, imply density profiles that are not particularly steep with respective indices of  $\alpha =$ 1.0, 1.2, and 1.3.
A steep $AMD$ over {\it narrow} ranges of $\xi$, on the other hand, is difficult to rule out. 
Indeed, by inspecting Fig.~\ref{fAMD}, it can be seen that some outflows (e.g., \ngc, \mcg, \ngcfifty) tentatively show steeper $AMD$s at low ionization ($\xi$) that flatten out at higher $\xi$.
By analyzing eq. (\ref{AMDxi}), a few special cases can be easily examined.

For a {\it constant-density} $n_0$ absorber, $\alpha$ = 0, and the $AMD$ would slowly decrease with ionization parameter  as 
$AMD(n_0) \propto \xi ^{-1/2}$.
No slope with $a<0$ is observed in the five present outflows, which rules out a constant density outflow. Indeed,  the constant density scenario is quite unphysical.

Yet another possible wind scenario is a steady mass conserving radial flow similar to a stellar wind, in which the mass outflow rate, opening angle, and wind velocity are all constant.
This results in a density profile $n \propto r^{-2}$, or $\alpha = 2$. Such a flow was assumed for instance by \citet{krolik95}.
In terms of the $AMD$ slope $a$, it tends to infinity (eq.~\ref{AMDxi}).
In this case, the ionization parameter would of course be constant ($\xi _0$) along the flow implying a sharply peaked $AMD$ (around $\xi _0$). 
This is in stark contradiction with the present sample that all feature broad $AMD$s.
It is worth noting however, that some Seyferts do show distinctly faster (few 1000~\kms) and more highly-ionized ($\log \xi > 3$) outflows that perhaps could be characterized by a single ionization parameter, a good example of which is the high-velocity components (4500~\kms\ \& 1900~\kms, respectively) found in NGC~4051 \citep{steenbrugge09} and in \mcg\ \citep{sako03}.
The low and intermediate charge states are clearly absent from these high-velocity components that do not have the same broad $AMD$ as the present, slower winds.  
This along with the lack of a continuous velocity distribution (c.f., distinct slow and fast components) suggests that the slow and fast outflows are different types of winds,
and may also differ in their position within the \agn\ system.

All of the outflows observed to date for which an $AMD$ could be reconstructed imply a flat to modestly {\it increasing} $AMD$ with $\xi$.  The column density measured for the low-ionization species is generally never higher than that of the high ionization plasma in terms of equivalent $N_H$.
In order to obtain a strictly flat $AMD$ while retaining the assumption of a global power-law density dependence on radius (eq. \ref{nalpha}), the density scaling must obey $n(r) \propto 1/r$ (or $\alpha = 1$).
Allowing for a moderately increasing $AMD$ with $\xi$, as observed in Fig.~\ref{fAMD} results in density scalings $r^{-\alpha}$ with $1 < \alpha < 1.3$. See Table~\ref{Tfit} for details.

\subsection{Small Scale Outflow}
\label{small}

Perhaps an equally interesting case is a distant absorber (at $r_0$) in which the ionization parameter distribution $\xi \approx L / (nr_0^2)$ is driven by density gradients on much smaller distances ($\delta r \ll r_0$).
The broad $AMD$ then requires the density to vary by several orders of magnitude to account for the different orders of magnitude of observed $\xi$.  
Such a multi-phase scenario could be a natural consequence, e.g., of turbulence as observed in the interstellar medium (ISM) gas in our galaxy.
Indeed, such a description for Seyfert outflows was proposed by \citet{chelouche05} for \ngc\ and further systematically explored for similar outflows in \citet{chelouche08}.

We now assume the density varies locally on these small length scales $\delta r$

\begin{equation}
n(\delta r) \propto \delta r ^{-\alpha} ,
\label{delta}
\end{equation}

\noindent which merely reflects typical density variations with distance that are assumed to be isotropic.
In this case, the dependence of the $AMD$ on $\alpha$ is slightly different than in Eqs. (\ref{xir} - \ref{AMDxi}), as $\xi \propto n^{-1} \propto \delta r^\alpha$, subsequently $\frac{d (\delta r)}{d \xi} \propto \xi ^{-\frac{\alpha - 1}{\alpha}}$, $dN_H = n(\delta r)\frac{d (\delta r)}{d \xi}d\xi \propto \xi ^{-\frac{2\alpha -1}{\alpha}} d\xi$, and the $AMD$ scales with $\xi$ as

\begin{equation}
AMD (r = r_0) = \left| \frac {dN_H}{d\log \xi} \right| 
\propto \xi \left| \frac {dN_H}{d\xi} \right| \propto \xi ^{-\frac{\alpha - 1}{\alpha}} \equiv \xi ^a 
\label{AMDxi2}
\end{equation}

\noindent where again $a$ is the $AMD$ slope and thus here $a (r = r_0) = -(\alpha -1)/\alpha$, and inversely

\begin{equation}
\alpha = \frac{1}{1+a} \pm \frac{\Delta a}{\left( 1+a \right)^2}
\label{ass}
\end{equation}

\noindent which can be compared with the large-scale flow relation given in eq. (\ref{alpha}).
Here again, $\Delta a$ is the standard error of the fitted power law slope $a$.
The resulting values for the five outflows considered here are listed in Table~\ref{Tfit}.
The functional forms of the density scaling $\alpha$ in eqs. (\ref{alpha}) and (\ref{ass}) are generally different.  
However, in the relevant regime of observed slopes $a \approx 0$ (and only there), the value of $\alpha$ implied by both expressions is similar and close to unity. 
We obtain $1 < \alpha\ < 1.3$ for the large scale flows and $0.7 < \alpha\ < 1$ for the small scale (constant $r_0$) flows. 
Far from $a = 0$, the two would be significantly different. 

It has been shown by \citet{goncalves06} that strong density gradients can occur naturally in an absorbing slab that is in pressure balance as
thermal instabilities \citep[suggested also by the discontinuous $AMD$s in][]{tomer07} result in strong temperature jumps that in turn cause strong density gradients.
\citet{chelouche05} propose a different model for \ngc\ of thermal launching of a wind with assumed strong density gradients on small length scales and invoke a density scaling similar to eq. (\ref{delta}).
They obtained a density profile with $\alpha = 0.8$ (denoted there as $-\beta$),
which is in very good agreement with the present results.
\citet{chelouche08} recently applied the same model to a sample of outflows.
However, the values quoted in \citet{chelouche08} are $\alpha \approx 1.3 \pm\ 0.6$,  systematically higher than what we find here, although still consistent given their large errors.
In any event, the present method provides much tighter constraints on $\alpha$ (Table~\ref{Tfit}).

\section{Discussion}
\label{discussion}

The  $AMD$ slope defined in Eqs. (\ref{AMDxi}) and (\ref{AMDxi2}) provides a novel way to obtain a self similar density profile of the outflow.
The slopes available for a handful of outflows are all quite similar and lie in the range of $0 < a < 0.4$.
The fact that $a$ is not much larger than zero reflects the broad and relatively flat ionization distribution of Seyfert outflows.
Along with the moderate velocities of a few 100~\kms, these are the only two model-independent recurring features of such flows.
In \S \ref{phys}, we quite generally differentiate between two possible physical flows that could produce this $AMD$ profile.
One is a large scale AGN outflow that would need to have a density profile of $n \propto r^{-\alpha}$ with 
 power-law indices of $1 < \alpha\ < 1.3$ in order to reproduce the observed $AMD$.
Importantly, a smooth radial mass conserving flow of $n \propto r^{-2}$, for example, that would correspond to a single $\xi$ or a sharply peaked $AMD$ are conclusively ruled out,
and so is a constant-density absorber ($a = -0.5$).
 Another possibility is a remote absorber with strong density gradients on local length scales $\delta r$ in which case $n \propto \delta r^{-\alpha}$ and $0.7 < \alpha\ < 1$.

If the outflow originates in the accretion disk, as believed by many authors, it is of the large scale type.
The derived density profiles can thus be compared directly with an entire family of self-similar solutions for hydromagnetic (MHD) disk winds \citep{contopoulos94}.  
To obtain an MHD wind with a slope of $\alpha =$ 1.5, for example, as in \citet{blandfordpayne} would require a {\it global} $AMD$ slope of $a = 1.0$ that is clearly excluded by the data.
Such a high index is not seen in any of the Seyfert outflows (Fig.~\ref{fAMD}, Table~\ref{Tfit}).
Models from \citet{contopoulos94}, on the other hand, with $1 < \alpha\ < 1.3$ are consistent with the observed $AMD$s (the present $\alpha$ corresponds in their notation to $3 - 2x$).
In fact, the specific self-similar solution with $\alpha = 1$ was preferred by \citet{konigl94} for its minimal magnetic energy.
It is interesting to note that a totally different approach using the temporal behavior of accreting sources  suggests indices of $\alpha = 1 - 1.5$ for the Compton scattering medium that produces the continuum X-ray spectrum \citep{kazanas97, hua97}.  
Consequently, a detailed MHD wind model has been developed by Fukumura et~al. (in preparation) with an $n \propto r^{-1}$ profile that by definition produces a flat $AMD$.
Note that density profiles that significantly deviate from $n \propto r^{-2}$ have been reported out to the narrow line region (NLR). \citet{bennert06} measured values of $\alpha = 1.46 \pm 0.2$ and $\alpha = 1.14 \pm 0.1$, respectively, for a sample of Seyfert~1 and Seyfert~2 NLRs.

For local density gradients, it is instructive to think of pressure-balanced clouds.
This type of models uses an absorbing slab to represent the \agn\ outflow, and with careful numerical treatment of the radiative transfer problem through the slab can compute its temperature, density, and ionization distributions \citep{goncalves06}. 
The astrophysical context of such "slabs" and how they are launched remains to be sorted out,
but models such as those of \citet{goncalves06} already stand out by virtue of their continuous and broad $AMD$, which is observed for Seyfert outflows.
In fact, with some fine tuning, these models can reproduce the majority of the measured ionic column densities, as well as the observed unstable region of the $AMD$.
On the other hand, they also predict other unstable regions that are not necessarily observed as such.
The numerical nature of these radiative transfer codes does not straightforwardly provide for an analytical power law density profile to be directly compared with eqs.~(\ref{delta}) and (\ref{ass}).

An alternative natural source of density gradients on many scales is turbulence,
such as that found in the interstellar medium (ISM) of the Milky Way. 
For a recent review of ISM observations and the relevant physical processes, see \citet{elmegreen04}. A general feature of the ISM is its power spectrum

\begin{equation}
P(k) \propto k^{-p}
\end{equation}

\noindent observed to be approximately a power law over five decades in wavenumber  $k = 1/\delta r$ from 10$^{-15} - 10^{-10}$~cm$^{-1}$ \citep{armstrong95}.
The standard definitions are used here for $P(k) \equiv \hat{n}(k)\hat{n}(k)^*$ where $\hat{n}(k)$ is the Fourier transform of an observable quantity (e.g., density) $n(\delta r)$, and $\hat{n}(k)^*$ is its complex conjugate.  
Typically, 3D ISM power spectra are consistent with, or slightly steeper than the Kolmogorov spectrum for non-magnetic, incompressible turbulence in which $p = 11/3$, as a result of energy cascading from large to small scales with no preferred intermediate scale  \citep[][]{kolmogorov41}.
Somewhat steeper power spectra ($p = 4$) can be expected from random shocks \citep{saffman71}.
While in a magnetic, compressible turbulent medium, density fluctuations can scale differently on different length scales \citep{lithwick01}, the prevalent spectrum over most scales is a Kolmogorov one.  
Hence, for the most part a value of $p = 3.8 \pm 0.2$ is a reasonable approximation for the ISM.

In order to obtain a density scaling akin to eq.~(\ref{delta}) for turbulent media, the mean squared density variations (variance) can be estimated from the volume 3D integral in Fourier space of the power spectrum (Parseval's theorem) over a cell the size of $1/\delta r$ 

\begin{equation}
\left< \left| \delta n^2 \right| \right>  = \int P(k) d^3k \propto \delta r ^{p-3}
\label{variance}
\end{equation}

\noindent The root mean square (rms) of the density fluctuations in a volume $\sim \delta r^3$ thus scales as 

\begin{equation}
\delta n_\mathrm{rms}(\delta r) = \sqrt{\left< \left| \delta n^2 \right| \right>} \propto \delta r ^{\frac{p-3}{2}}
\label{rms}
\end{equation}

\noindent  If this rms of density fluctuations is understood as the typical scaling of density variations over $\delta r$ as in $n(\delta r) \propto \delta r ^{-\alpha}$ with $\alpha = (3-p)/2$, it can be directly compared with eq.~(\ref{delta}). 
Note that since generally $p > 3$, the density fluctuations do not vanish at any scale within the medium (i.e., no asymptotic homogeneity).
Thus, eq.~(\ref{rms}), in principle, could be meaningful over many orders of magnitude. 

Applying the typical ISM value of $p = 3.8 \pm\ 0.2$ yields a density scaling of $n \propto \delta r^{0.4}$ ($\alpha = -0.4 \pm\ 0.1$). 
Recall that using the $AMD$ slopes, we obtained markedly different density gradients with $ 0.7< \alpha\ < 1.0$  (Table~\ref{Tfit}).
Another difference between the Seyfert outflow and typical ISM is the range in density.
For a distant, localized ($r_0$) Seyfert absorber, the range in density required for producing the observed ionization structure needs to span up to five orders of magnitude ($-1 < \log \xi < 4$).
It is not clear that the ISM spans such a broad density range, and certainly not within a single medium (e.g., a molecular cloud). 
The observed five orders of magnitude in $k$ in the ISM correspond to only ($n \propto k^{-0.4}$) two orders of magnitude in density.
In order to obtain the high density gradients of $0.7 < \alpha < 1.0$,
an extremely shallow power spectrum with $p = 0.8 - 1.0$ would be required.
We are not aware of any other (ISM) medium with such a flat power spectrum.



\section{CONCLUSIONS}
\label{sec:concl}

We have presented the Absorption Measure Distribution ($AMD$) of five nearby AGN outflows.
The $AMD$ is the differential column density through the flow as a function of ionization parameter $\xi$
that can be derived from the high resolution X-ray absorption spectrum of the outflow.
Key to reconstructing the $AMD$s is the observation of almost all Fe charge states from neutral to H-like.
Using a power law approximation, we fit for the $AMD$ slopes and find that there is only a mild
increase of column density with $\xi$, namely the slopes are all in the range of 0 to 0.4. 
It may be possible that at low ionization $\log \xi < 1$, the $AMD$ is steeper and then flattens out towards higher ionization ($\log \xi > 2$).
However, the limited ionization resolution in the $AMD$ (due to the large formation regions of the various charge states) does not allow for a more precise assessment.  
In any event, the $AMD$ reconstruction at low ionization is currently hampered by the uncertain dielectronic recombination rates and by the unknown EUV continuum.
We expect the former limitation to be remedied soon by new laboratory measurements and improved atomic calculations. 
We also expect more high-quality grating spectra of \agn\ outflows to be released to the \chandra\ and \xmm\ archives in the next few years.
Future high throughput \x\ spectrometers, such as the SXS currently being built for \astroh\ will further increase the sample.
As the sample of well studied Seyfert outflows increases, the present method will continue to facilitate our growing understanding of the physics governing these winds.

We have discussed the best-fit $AMD$ slopes in the context of two qualitatively different scenarios, one is a continuous large-scale radial flow and the other is a distant well localized absorber.
If the outflow is of the former type, we can constrain its density profile rather tightly to $n \propto r^{-\alpha}$ with $1 < \alpha < 1.3$. 
This already rules out the standardly assumed $n \propto r^{-2}$ profile, or even a profile of $n \propto r^{-1.5}$. 
A constant-density absorber ($\alpha = 0$) is also clearly ruled out.
An MHD wind, on the other hand, with a shallower density profile of $\alpha \gsim 1$ as proposed by Fukumura et al. (in preparation) is consistent with the observed outflows.
Alternatively, slab models with proper radiative transfer can produce a broad ionization distribution, 
with intermediate thermally unstable regions, generally consistent with the observed $AMD$s. 
If the absorber is remote from the central source, and compact, then its ionization depends solely on the local density gradients.
We find that in this case, the density should vary as $n \propto \delta r^{-\alpha}$ with $0.7 < \alpha < 1$.
Such sharp gradients, and occurring over five orders of magnitude in density, are quite different from those typically found in the Milky Way ISM and interpreted as due to turbulence.

\acknowledgments
I thank Tomer Holczer for providing the $AMD$ results in simple data form and Alex Blustin for suggesting the title for the paper.

\clearpage

\begin{deluxetable}{lccc}
\tabletypesize{ \footnotesize }
\tablecolumns{4} \tablewidth{0pt}
\tablecaption{Best-fit parameters from the linear fits in Fig.~\ref{fAMD}
$AMD (\xi) \propto \xi ^a$}
\tablehead{
   \colhead{Target} &
   \colhead{Slope (a)} &
   \colhead{$\alpha$ \tablenotemark{a}} &
   \colhead{$\alpha$ \tablenotemark{b}} 
}

\startdata

\ngc & 0.29 $\pm$ 0.11 & 1.22 $\pm$ 0.07 & 0.78 $\pm$ 0.07 \\
\iras & 0.02 $\pm$ 0.01 & 1.02 $\pm$ 0.01 & 0.98 $\pm$ 0.01\\
\mcg & 0.10 $\pm$ 0.11 & 1.10 $\pm$ 0.09 & 0.91 $\pm$ 0.09\\
\ngcfifty & 0.14 $\pm$ 0.08 & 1.12 $\pm$ 0.06 & 0.88 $\pm$ 0.06\\
\ngcseven & 0.24 $\pm$ 0.13 & 1.19 $\pm$ 0.08 & 0.80 $\pm$ 0.08\\
\enddata
\footnotesize
\tablenotetext{a}{For a distance driven $AMD$, $\alpha = (1+2a)/(1+a)$, as in \S \ref{large}}
\tablenotetext{b}{For a density driven $AMD$, $\alpha = 1/(1+a)$, as in \S \ref{small}}

\label{Tfit}
\end{deluxetable}

\clearpage

\end{document}